%% file: hse-test.tex
\documentclass[journal=jpclat,manuscript=article]{achemso}

\usepackage{graphicx}
\usepackage{float}
\usepackage{subfig}
\usepackage{chemformula}
\usepackage[T1]{fontenc}
\usepackage{amsmath,amssymb}
\usepackage{dcolumn}
\usepackage{bm}
\usepackage{epsfig}
\usepackage{longtable}
\usepackage{multirow}
\usepackage{color}
\usepackage{hyperref}

\include{newcommands}

\author{Yuyang Ji}
\affiliation{CAS Key Laboratory of Quantum Information, University of Science and Technology of China, Hefei 230026, Anhui, China}
\author{Peize Lin}
\affiliation{Songshan Lake Materials Laboratory, Dongguan 523808, Guangdong, China}
\email{linpeize@sslab.org.cn}
\author{Xinguo Ren}
\email{renxg@iphy.ac.cn}
\affiliation{Songshan Lake Materials Laboratory, Dongguan 523808, Guangdong, China}
\altaffiliation{Institute of Physics, Chinese Academy of Sciences, Beijing 100190, China}
\author{Lixin He}
\email{helx@ustc.edu.cn}
\affiliation{CAS Key Laboratory of Quantum Information, University of Science and Technology of China, Hefei 230026, Anhui, China}

\title{Reproducibility of Hybrid Density Functional Calculations for Equation-of-State Properties and Band Gaps}

\begin{document}

\begin{tocentry}
 \centering
 \includegraphics[width=3.25in,height=1.75in]{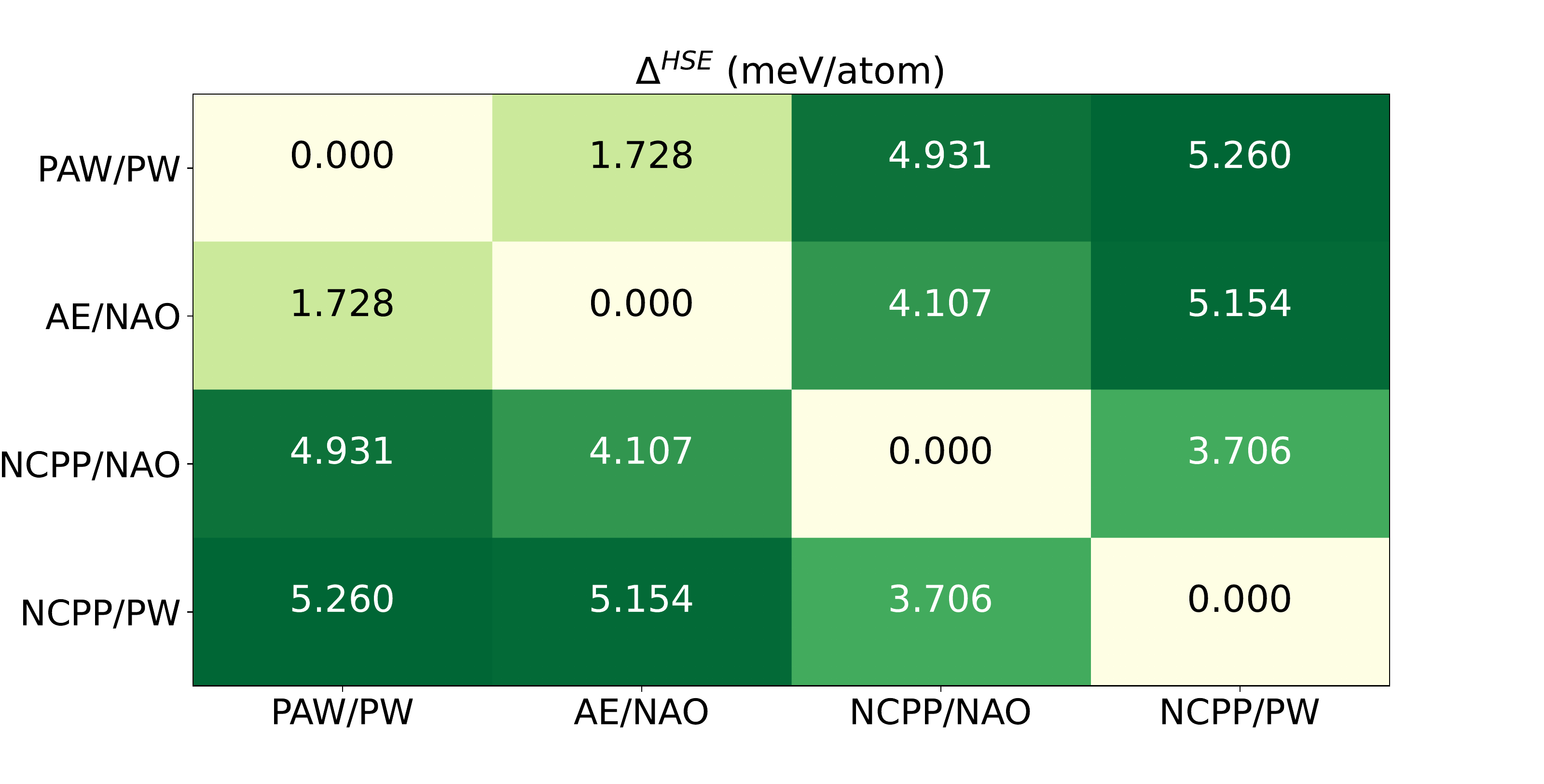}
\end{tocentry}

\begin{abstract}
Hybrid density functional (HDF) approximations usually deliver higher accuracy than local and semilocal approximations to the exchange-correlation
functional, but this comes with drastically increased computational cost.
Practical implementations of HDFs inevitably involve numerical approximations -- even more so than their local and semilocal counterparts due
to the additional numerical complexity arising from treating the exact-exchange component.
This raises the question regarding the reproducibility of the HDF results yielded by different implementations. In this work, we benchmark
the numerical precision of four independent implementations of the popular Heyd-Scuseria-Ernzerhof (HSE) range-separated HDF
on describing key materials' properties, including both properties derived from equations of states (EOS) and band gaps of 20 crystalline solids.
We find that the energy band gaps obtained by the four codes
agree with each other rather satisfactorily. However, for lattice constants and bulk moduli,
the deviations between the results computed by different codes are 
of the same order of magnitude as the deviations between the computational and experimental results. 
On the one hand, this means that the HSE functional is rather accurate for describing the
cohesive properties of simple insulating solids. 
On the other hand, this also suggests that 
the numerical precision achieved with current major HSE implementation is not sufficiently high to unambiguously assess the physical accuracy of HDFs.    
It is found that the pseudopotential treatment of the core electrons is a major factor that contributes to this uncertainty. 

\end{abstract}

\section{Introduction}
The reproducibility of the results, both experimental and computational ones, is essential for scientific researches. 
In computational materials science which heavily relies 
on computer simulations, this is particularly a concern. Ideally, one would like to see that the 
same results be obtained regardless which computer code is used. As one of the most widely used first-principles approaches 
for simulating materials' properties, the generalized gradient approximations (GGAs) in density functional theory (DFT), such as the Perdew-Burke-Ernzerhof (PBE) \cite{perdew1996generalized} functional, have been implemented
in a plethora of computer codes and widely applied in condensed physics, quantum chemistry, and materials science. A natural question arises regarding the reproducibility of the
GGA results as obtained by different computer codes. This question was addressed by Lejaeghere \textit{et al.} in Ref.~\citenum{Lejaeghere.2016} where the reproducibility of
the PBE implementations has been examined over 15 different computer codes, using 40 different potentials and/or basis set types,  
for describing the equation of states (EOS) of 71 elemental crystals. These authors introduced
a so-called $\Delta$ gauge - a single value that measures the deviations of the EOS curves yielded by two different codes -- and found that the $\Delta$ values
between mainstream DFT codes, averaged over the elemental crystals, are at the level of
1 meV. Such a remarkable numerical precision assures the reproducibility of the mainstream DFT codes, as far as the PBE functional concerned.

In recent years, the hybrid density functionals (HDFs) are getting more and more popular in computational materials science.
By incorporating a fraction of Hartree-Fock exchange (HFX), HDFs successfully addressed one big issue of 
local-density approximation (LDA) and GGA-type functionals -- the severe underestimation of the band gaps of semiconductors and insulators. 
Popular HDFs include B3LYP\cite{Becke1993, stephens1994ab}, PBE0\cite{perdew1996rationale} 
and HSE\cite{heyd2003hybrid, Heyd.2006, krukau2006influence}. Among these, the HSE functional belongs to the screened HDFs in which 
the Coulomb interaction is split into short-range and long-range components and
only the HFX based on the short-range Coulomb interaction is mixed into the exchange-correlation (XC) functionals. 
Incorporating the HFX only for the short-range part of the Coulomb interaction is arguably better
suited for describing narrow- and middle-gap semiconductors than global HDFs \cite{heyd2003hybrid}.
In practice, the HSE functional shows remarkable performance not only for band gaps,
but also for structural and cohesive properties\cite{Heyd.2005}.
The price to pay, however, is the significantly increased computational cost, arising from the evaluation of the short-range HFX, compared to GGAs.

Nowadays, HDFs have been implemented in various mainstream first-principles code packages, employing different potentials, basis sets, and numerical
approximations. For example, 
in the Vienna Ab initio Simulation Package (VASP) \cite{kresse1993ab, kresse1996efficient}, the HFX is computed within the projector augmented wave (PAW) formalism and plane-wave (PW) basis set \cite{paier2005perdew, Paier2006, Paier2006b}. The VASP implementation reduces the computational cost by downsampling the $\bfk$-point grid in the HFX evaluations
\cite{Paier2006}. As for atomic-orbital based codes, the resolution-of-the-identity (RI) \cite{Whitten.1973, Dunlap.1979, FEYEREISEN1993359, VAHTRAS1993514, WEIGEND1998143, Brett.2010, ren2012resolution, ihrig2015accurate, levchenko2015hybrid} is often employed to reduce the computational cost of HFX component. 
In particular, using the localized variant of the RI approximation, one can achieve a linear scaling of the computational cost for building the HFX matrix
 with respect to the system size $N$.  \cite{levchenko2015hybrid,lin2020accuracy,Lin2021}
The localized RI (LRI) based HDFs have been implemented within the Fritz Haber Institute \textit{ab initio} molecular simulations (FHI-aims) code package \cite{blum2009ab} and 
recently also in the atomic-orbital-based \textit{ab initio} computation at UStc (ABACUS) code package \cite{chen2010systematically, li2016large}. 
Besides these efforts, Lin developed an adaptively compressed exchange operator (ACE) \cite{lin2016adaptively} scheme, which, together with  a two-level self-consistent field
iterator procedure, can significantly  accelerate the HDF calculations using PW basis sets. Such a scheme has been implemented in the
DGDFT\cite{lin2012adaptive, hu2015dgdft} and the Quantum ESPRESSO (QE)\cite{giannozzi2009quantum, giannozzi2017advanced} code packages.
Other techniques to speed up the HFX calculations include the exploration of the localization offered by Wannier representation \cite{wu2009order,Dziedzic.2013,Dziedzic.2021}, 
intepolative separable density fitting approaches\cite{Lu/Yin:2015,Qin.2020}, and representing NAOs in terms of Gaussian-type orbitals \cite{Shang/Li/Yang:2010}, etc.

As alluded to above, the numerical precision and the reproducibility of PBE implementations in mainstream DFT codes have been established via the $\Delta$-value project \cite{Lejaeghere.2014, Lejaeghere.2016}. However, this conclusion may not be automatically extended to HDFs as implemented in these codes. 
In fact, the numerical precision of HDF calculations does not only rely on the treatment
of electron-ion interactions (pseudopotentials vs. all-electron) and basis set types (e.g., plane waves vs. atomic orbitals), but also on the numerical approximations
involved in the evaluation of the HFX potential and energy. To our knowledge, a systematic assessment of the numerical precision of different
HDF implementations has not yet been reported.   
In this work, utilizing the HSE functional available in four computer codes, including VASP, FHI-aims, QE, and ABACUS,  whose HDF
implementations differ considerably in algorithmic and numerical details, we assess the reproducibility of HSE results 
by computing the pairwise $\Delta$ values among these
four codes, based on the energy-volume curves of 20  binary compounds. These 20 compounds have been widely used as a test set 
for evaluating the performance of HDFs
 \cite{Heyd.2004, Heyd.2005, Garza.2016, Cui.2018, Zhang2018b, lin2020accuracy}. Such a $\Delta$-value based assessment will
be corroborated by additional properties including the lattice constants, bulk moduli, 
and band gaps.  Naturally, benchmarking the numerical precision of
HDF implementations over all mainstream DFT codes and across different types of materials requires community-wide efforts and 
goes beyond the scope of the present paper.
Nevertheless our work provides a solid basis that will hopefully stimulate more benchmark efforts in this direction.

\section{Methods}
In this section, we recapitulate the essentials of the numerical frameworks and computational setups employed in the four codes considered in this work. 
First of all, different strategies are used to treat the effects arising from the nuclei and core electrons:
The VASP code employs the PAW method 
whereas QE and ABACUS make use of the  SG15 optimized norm-conserving Vanderbilt-type (ONCV) pseudopotentials \cite{vanderbilt1990soft, schlipf2015optimization}. In contrast, FHI-aims treats all electrons on an equal footing. Scalar relativistic effects are included either in
the PAW or norm-conversing pseudopotentials (NCPP), or via the ``atomic ZORA'' scheme \cite{Lenthe.1996, Huhn.2017} in the all-electron case.
The spin-orbit coupling effect is not included in the HSE calculations here.

Regarding the basis sets, VASP and
QE both use PWs to expand the Bloch wavefunctions, whereas numerical atomic orbital (NAO) basis sets are used in the FHI-aims and ABACUS codes, although
different NAO construction strategies are adopted in these two codes, in order to best represent their respective all-electron (AE) or pseudo-wavefunctions. 
Moreover, kinetic energy cutoffs of $550$ eV and $100$ Rydberg (Ry) are used for the PW expansion of wavefunctions in VASP and QE calculations, 
respectively. 
For FHI-aims, the default ``\emph{tight}'' NAO basis sets \cite{blum2009ab} are adopted, whereas  the so-called ``DZP-DPSI'' \cite{lin2021strategy} basis sets are employed in ABACUS. 
Additionally, $100$~Ry energy cutoff is used in ABACUS for the Hartree potential calculation and for determining the uniform quadrature grid. In Table~\ref{tab:settings}, we summarize
the computational settings for the four codes, together with the convergence thresholds in the self-consistent calculations. Please note that
these settings are roughly the default ones that are typically used in production calculations.  Our intention here is not to push towards
the limit of high-precision settings of each code, but rather to check the numerical precision that is achieved in routine calculations. 
An investigation of the influence of difficult levels of basis sets and numerical settings on the achieved precision of the results has recently
been reported by Carbogno et al. \cite{Carbogno.2022} for three different computer codes, including FHI-aims.

As mentioned above, additional complexity is involved in the evaluation of the HFX component of the HDFs. For VASP, the standard HSE implementation as available
in VASP 5.4.1 is used. As for QE calculations,  we use its PWSCF module v.6.5 where the ACE technique \cite{lin2016adaptively} is used 
as the default to calculate the HFX contributions\cite{giannozzi2017advanced}. Finally, for NAO-based codes, the LRI approximation is used in both FHI-aims and ABACUS for
evaluating the HFX term. To render the numerical errors associated with LRI sufficiently small, 
an enhanced auxiliary basis set has to constructed, compared to
conventional global Coulomb-metric based RI scheme \cite{ihrig2015accurate}. In FHI-aims, this is achieved by adding a hydrogen-like 5 $g$ function with
an effective nuclear charge $Z=6$ for each element  that is solely used to generate extra auxiliary basis functions (ABFs) \cite{ihrig2015accurate}. 
In ABACUS, additional optimized ABFs can be included to complement the ``on-site'' ones to increase the numerical accuracy \cite{lin2020accuracy}.
Furthermore, an efficient pre-screening procedure is invoked to filter out negligibly small contributions to achieve a linear-scaling build of
the HFX. The settings of the ABF basis sets, and the pre-screening parameters for ABACUS HSE calculations are given in the Supporting Information (SI)
(cf. Table~S1 and discussions therein).

%


The equilibrium volume $V_0$ and bulk modulus $B_0$ of each solid are determined via the Birch–Murnaghan EOS \cite{birch1947finite}. Since our 
test set of materials only comprises fcc crystals, the lattice constants $a_0$ is found from the equilibrium volume $V_0$ via 
$a_0= \left(4V_0\right)^{1/3}$. The $\Delta$ value for a material $\alpha$ between any two codes are obtained from the calculated EOS curves, following
the procedure proposed in Ref.~\citenum{Lejaeghere.2014}. 
For a more trustworthy comparison between experimental cohesive properties and the HSE predictions, 
the zero-point motion (ZPM) effects are taken into account, utilizing the values as estimated in Ref.~\citenum{Zhang2018b} for all calculations. Finally, the band gaps are estimated on a discrete $ 8\times8\times8$ $\bfk$-point mesh, and the experimental lattice parameters are used for all the band structure calculations.

\begin{table}[H]
    \centering
    \caption{Numerical frameworks and computational settings for the four codes included in the present benchmark studies. ``Method'' describes how the effects from the nuclei and core electrons are treated. ``Basis sets'' tell the type of basis functions to expand the wavefunctions. $E_\text{cutoff}$: the kinetic energy cutoffs for the PW expansions (in case of VASP and QE), and for the Hartree potential calculation and for determining the uniform quadrature grid (in case of ABACUS).
    $\eta_{etot}$, $\eta_\rho$ denotes the convergence threshold for the self-consistency cycle, based on the criteria of total energy, charge density, respectively. (n/a: this metric is not used in the particular code or not set in this benchmark test).}
    \label{tab:settings}
    \begin{tabular}{ c|c|c|c|c|c }
         \hline
                  &  Method &  Basis sets      &  $E_\text{cutoff}$    &  $\rm \eta_{etot}$     &  $\rm \eta_{\rho}$      \\
         \hline
         \hline
         VASP     &  PAW    &  PW              &  550 (eV)    &  $ 10^{-6}$ (eV)  &  n/a                            \\
         FHI-aims &  AE     &   NAO (\textit{tight})    &  n/a          &  n/a  &  $10^{-6}$ \\
         QE       &  $\rm NCPP^a$   &  PW              &  100 (Ry)    &  $10^{-6}$ (a.u) &  n/a                            \\
         ABACUS   &  $\rm NCPP^a$   &  NAO (DZP-DPSI)     &  100 (Ry)    &  n/a                  &  $10^{-8}$               \\
         \hline
         \hline
    \end{tabular}
\end{table}
a: the scalar-relativistic SG15 optimized norm-conserving Vanderbilt-type (ONCV) pseudopotentials\cite{vanderbilt1990soft, schlipf2015optimization}



\section{Results}
\subsection{Equation of states}

\begin{figure}[htbp]
    \centering
    \includegraphics[scale=0.65]{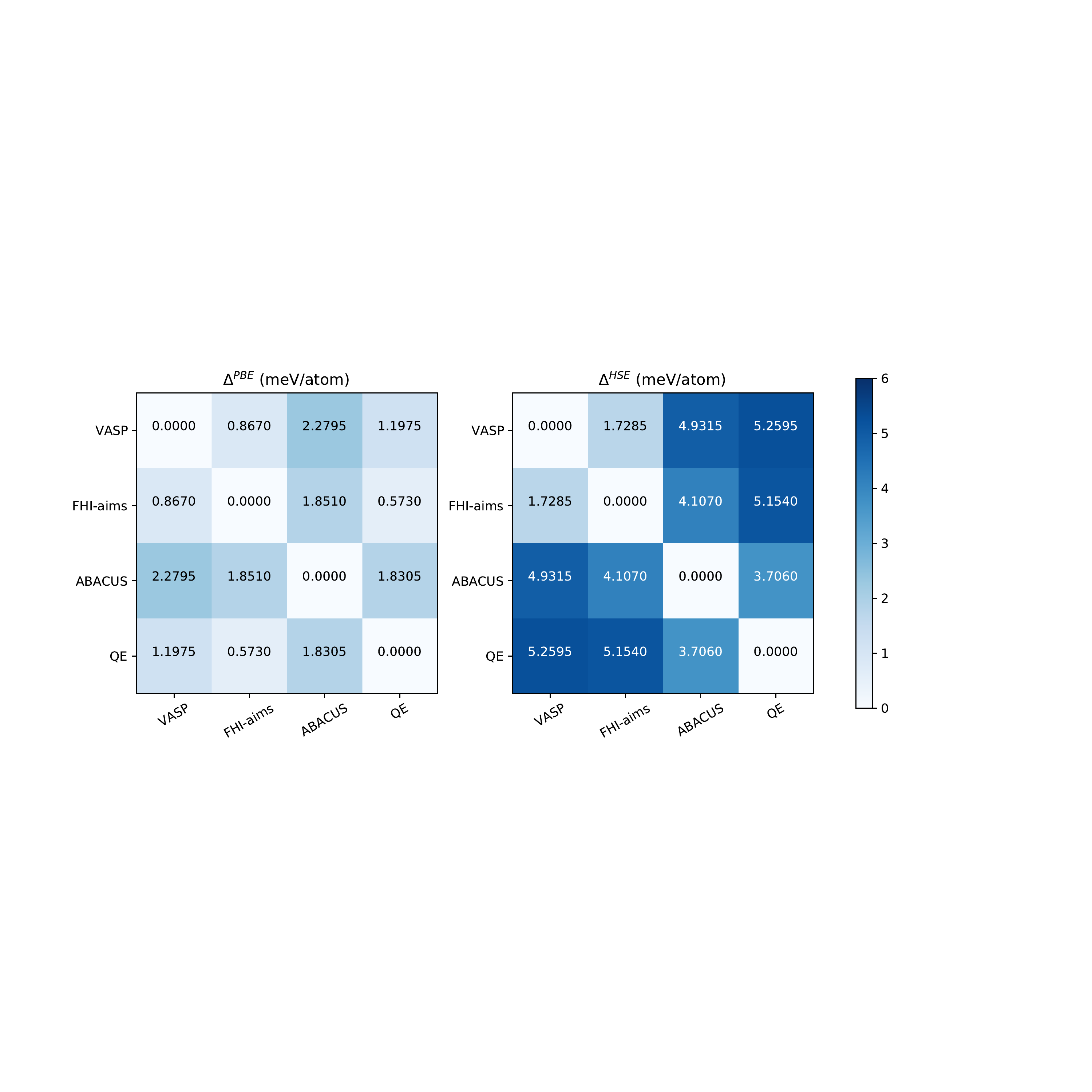}
    \caption{The the $\Delta$-value matrix (in meV/atom) that measures the pair-wise deviations between the EOS curves generated by four codes, i.e., 
    VASP, FHI-aims, ABACUS, and QE for both PBE (left) and HSE (right) functionals. }\label{fig:Delta_matrix}
\end{figure}

We start by examining the pairwise $\Delta$ values evaluated in between the four codes, given by an average of the individual $\Delta$ values for the 20 
materials. Namely,
\begin{equation}
    \Delta_{ij} = \frac{1}{20} \sum_{\alpha = 1}^{20} \Delta_{ij}(\alpha)
    \label{eq:Delta_ab}
\end{equation}
where 
\begin{equation}
    \Delta_{ij}(\alpha) = \sqrt{\frac{\int_{0.94V_{0,\alpha}}^{1.06V_{0,\alpha}}{\left[E_{i,\alpha}(V)-E_{j,\alpha}(V)\right]^2dV}}{0.12V_{0,\alpha}}}
    \label{eq:Delta_ij_alpha}
\end{equation}
measures the root-mean-square difference between the energy-volume ($E$-$V$) curves (with aligned energy minima) generated by two codes $i$ and $j$ for a material $\alpha$ in the test set.  
In Eq.~\ref{eq:Delta_ij_alpha},  $V_{0,\alpha}$ is the averaged equilibrium volume of the $\alpha$-th material, as predicted by the two codes. 
A change of the volume within a range of
$-6\%$ to $6\%$ around the equilibrium point is included in the evaluation of the $\Delta$ value. The thus determined $4\times 4$ $\Delta$ matrix is shown in Figure~\ref{fig:Delta_matrix} for both the PBE and HSE functionals.
From Figure~\ref{fig:Delta_matrix}, one can immediately see that the $\Delta$ values 
for the HSE functional (right panel) are 2 to 3 times larger than their counterparts for the PBE functional (left panel).
This is not entirely surprising since, as mentioned above, additional numerical approximations are involved in the evaluation of the HFX component
of the HSE functional and this will very likely yield additional numerical discrepancies between different implementations. 
Among the four codes, the smallest $\Delta$ value for the HSE functional, 1.7 meV/atom, is obtained between the all-electron FHI-aims code and 
the PAW-based VASP code,  whereas the 
largest  $\Delta$ value, 5.3 meV/atom, occurs between VASP and QE. Although VASP and QE both use the PW basis sets, the norm-conserving pseudopotentials 
(NCPPs) are used in QE whereas the PAW scheme is used in VASP. The $\Delta$ matrix in Figure~\ref{fig:Delta_matrix} suggests that 
the approach to describe 
the valence-ion interactions, has a bigger influence than the choice of basis sets. Indeed, a $\Delta$ value of 3.7 meV/atom is obtained between the two NCPP-based codes -- QE and ABACUS, which
is smaller than the discrepancy between the two PW-based codes (5.3 meV/atom) and between the two NAO-based codes (4.9 meV/atom). Comparing the two
panels in Figure~\ref{fig:Delta_matrix}, one can see that the 
distribution pattern of the $\Delta$ values among the four codes are rather similar for the PBE and HSE functionals, except that the $\Delta$ values are 
2-3 times larger for the HSE functionals, and that
the discrepancy between VASP and QE results is less pronounced for the PBE functional.   
The reason that the two PW-based codes show larger deviations for the HSE functional, while it is much less so for the PBE functional,
is probably because the PAWs and NCPPs used in the present calculations are constructed for the PBE functional and not for the HSE functional.
It is not fully clear yet whether the ACE approach, that is used in QE calculations, plays any role here.


\begin{figure}[htbp]
    \subfloat[\label{fig:lattice_a}]{\includegraphics[scale=0.45]{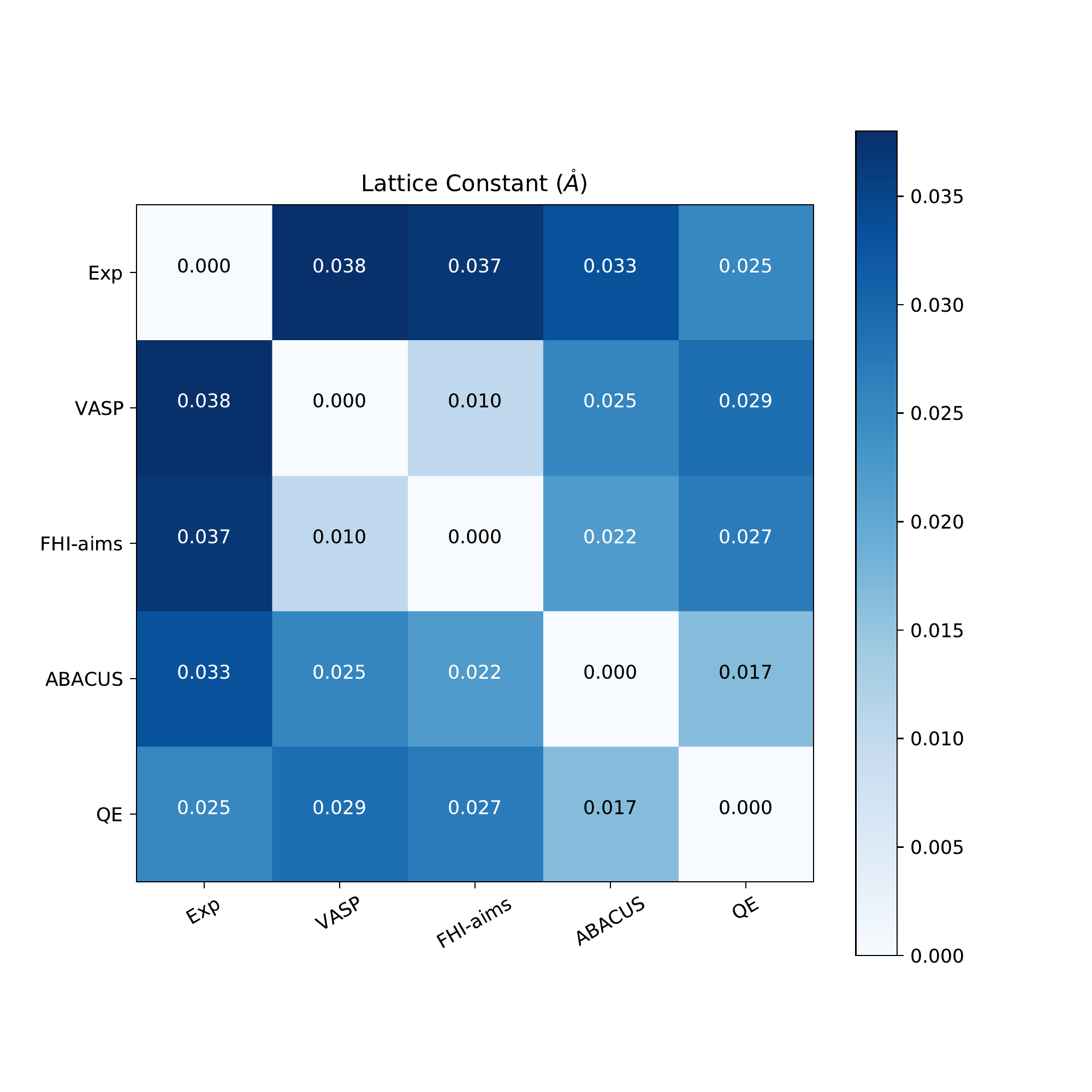}}
    \\
    \subfloat[\label{fig:lattice_b}]{\includegraphics[scale=0.45]{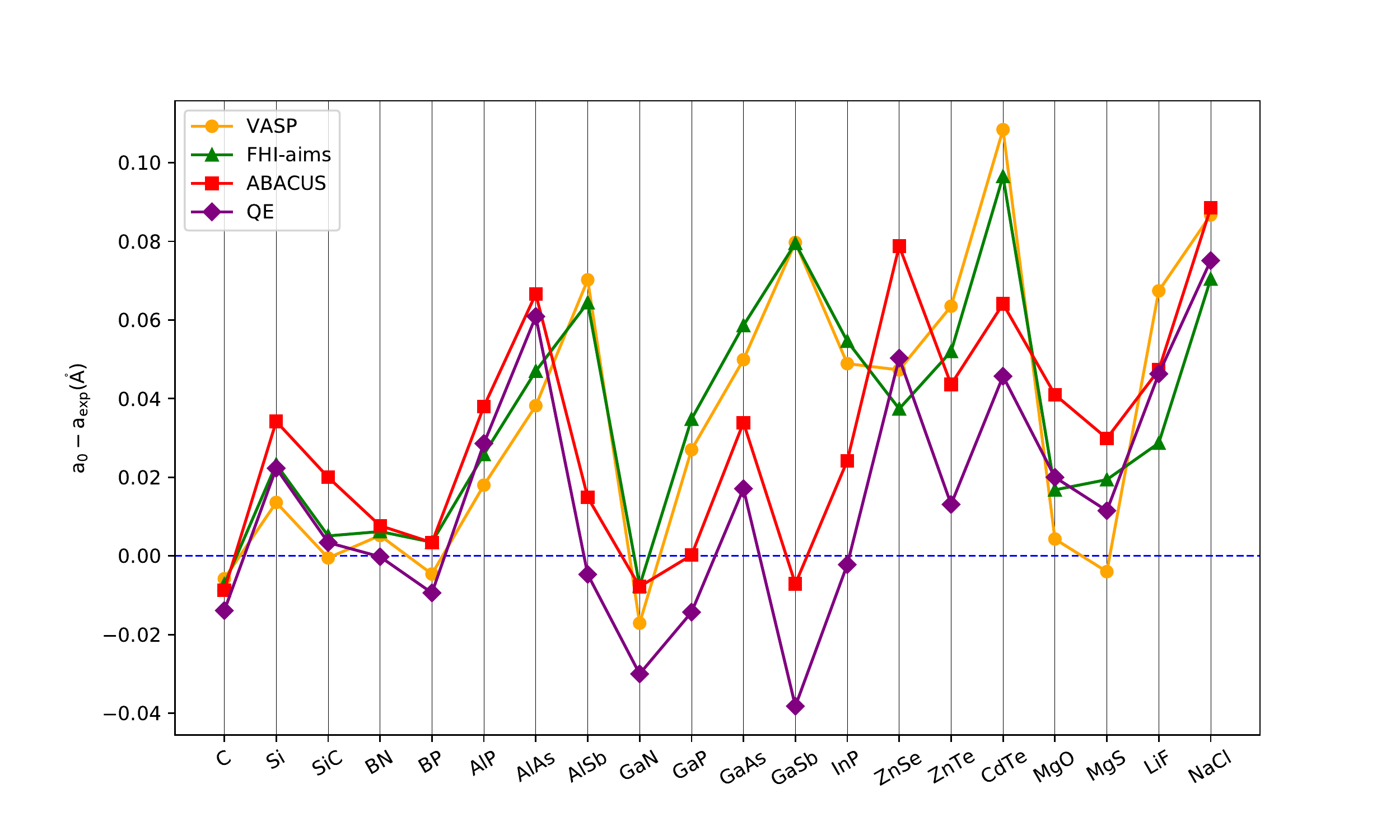}}
    \caption{(a) The pairwise MADs between the computed HSE lattice constants (in \AA ) by VASP, FHI-aims, ABACUS, QE, and between
    the computational and experimental values. 
    The colormap is such that the darker the color, the greater the MADs.
    (b) Deviations between the four sets of HSE lattice constants and the experimental value for individual
    compounds. The computational lattice constants are corrected for the ZPM effects in both (a) and (b).} 
\end{figure}

Next we check how the lattice constants determined by the four HSE implementations agree with one another. In this case, 
the experimental values are also included for comparison. 
In Figure~\ref{fig:lattice_a}, we present the pairwise mean absolute deviations (MADs) between the four sets of
computational HSE lattice constants, as well as those between the computational and experimental results. Specifically, what
is presented in Figure~\ref{fig:lattice_a} are
\begin{equation}
{\rm MAD}(a_0,i,j)=\frac{1}{20} \sum_{\alpha=1}^{20} \left| a_{0,i}(\alpha)-a_{0,j}(\alpha)\right|
\end{equation}
where 
$a_{0,i}(\alpha)$ is the equilibrium lattice constant for the $\alpha$-th material as determined by
experiment ($i=1$) or by one of the four HSE implementations ($2\le i \le 5$). Here, 
the computational lattice constants are corrected for the ZPM effect. For simplicity, we have assumed that the ZPM corrections are not sensitive to
the HSE implementations, and the same ZPM correction values for the HSE lattice constants, as reported in from Ref.~\citenum{Zhang2018b}
(obtained using FHI-aims),
are used here for all the computed results. As such, the MADs between the computational results are not affected by the ZPM corrections,
and only those between experimental and computational ones are.
From Figure~\ref{fig:lattice_a},
one can clearly see that the computational results fall into two groups, i.e., VASP and FHI-aims in one group, and
ABACUS and QE in another, with the MADs between the codes within the same group about a factor of 2 smaller than those 
between the two groups. This again shows, consistent with the $\Delta$ test, that the all-electron and PAW results agree best, whereas
larger deviations of the all-electron/PAW results from the NCPP-based results are observed. As far as the all-electron/PAW results 
concerned, the deviations between the calculated results are appreciably smaller than the deviations between the computational and
experimental results. However, this does not hold any more when all four sets of computational results are taken into account.
Finally, one can see that VASP and FHI-aims
results show larger deviations from the experimental results than the ABACUS and QE results do. This probably comes from fortuitous 
error cancellations.

To have a closer look at what is happening behind, in Figure~\ref{fig:lattice_b} we further present the
deviations of the computed lattice constants from the experimental values for the 20 individual compounds.  
The actual values of the computational and experimental lattice constants are further presented in Table~S2 of SI.
Figure~\ref{fig:lattice_b} shows the VASP and FHI-aims results follow each other closely, whereas QE and ABACUS
results show a similar zigzag pattern as the deviations vary among different materials. Such behavior is consistent
with the pairwise MAD analysis presented in Figure~\ref{fig:lattice_a}, but Figure~\ref{fig:lattice_b} also reveals
important details. Although for a large part of materials, the four codes yield similar HSE lattice constants, but exceptions do exist.
For example, for GaSb, FHI-aims and VASP predict that HSE overestimates the lattice constant by 0.08 \AA, whereas QE
suggests there is an underestimation of 0.04 \AA. Finally, ABACUS yields a HSE lattice constant there is in excellent, but most
likely fortuitous agreement with experiment. Similar phenomenon is observed for AlSb whereby VASP and FHI-aims still
show a pronounced overestimation, but in this case the HSE lattice constant
from QE is in best agreement with experiment, whereas ABACUS shows a slight overestimation. This benchmark test tells that,
for some materials, the current HSE implementations may yield appreciably different results than may change the direction
of the error when compared to experiment. The way to describe the interactions between valence electrons
and ion cores (all-electron, PAW, or NCPPs) seems to play a major role in affecting the outcomes of HSE calculations.


\begin{figure}[htbp]
    \centering
    \subfloat[\label{fig:bulkmoduli_a}]{\includegraphics[scale=0.45]{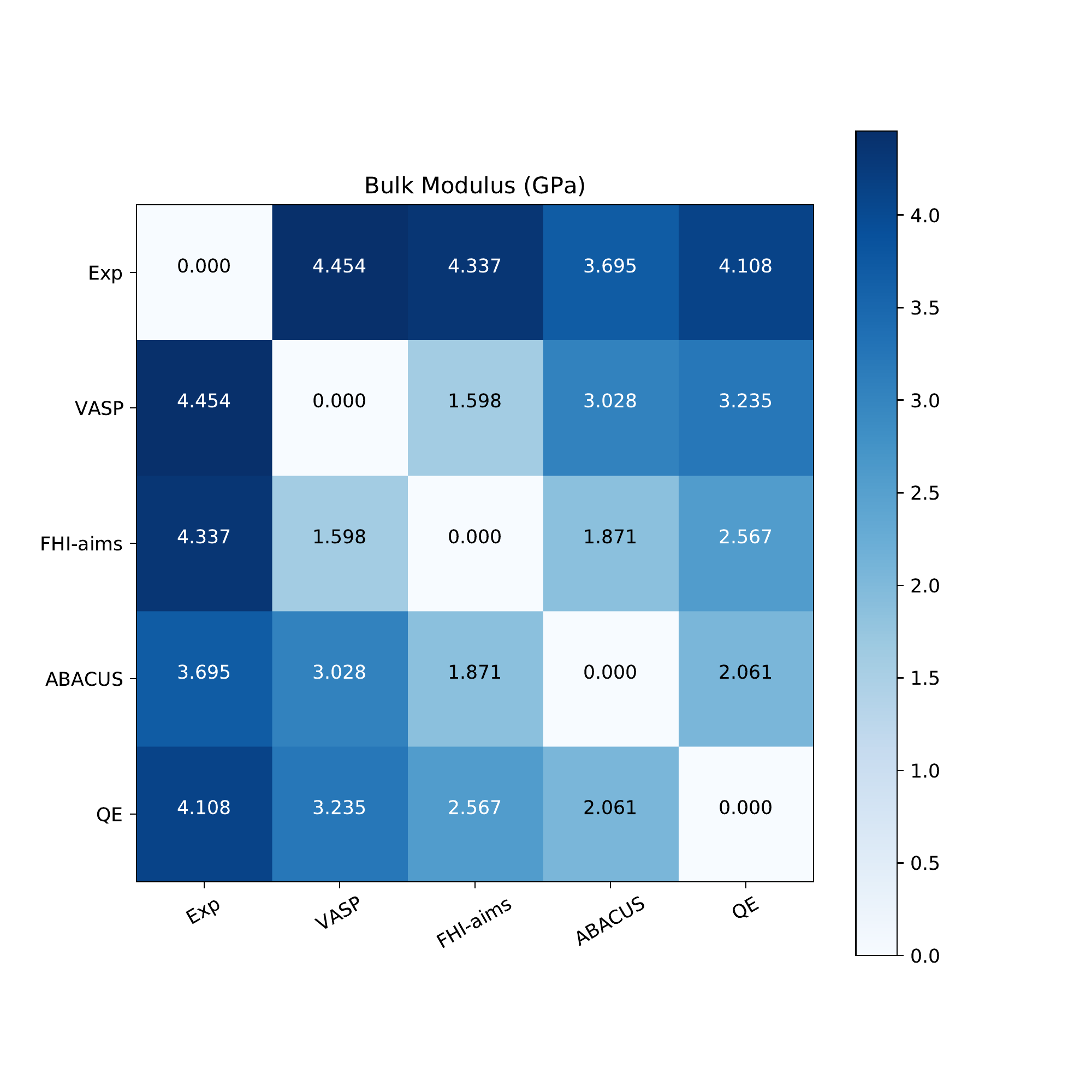}}
    \\
    \subfloat[\label{fig:bulkmoduli_b}]{\includegraphics[scale=0.45]{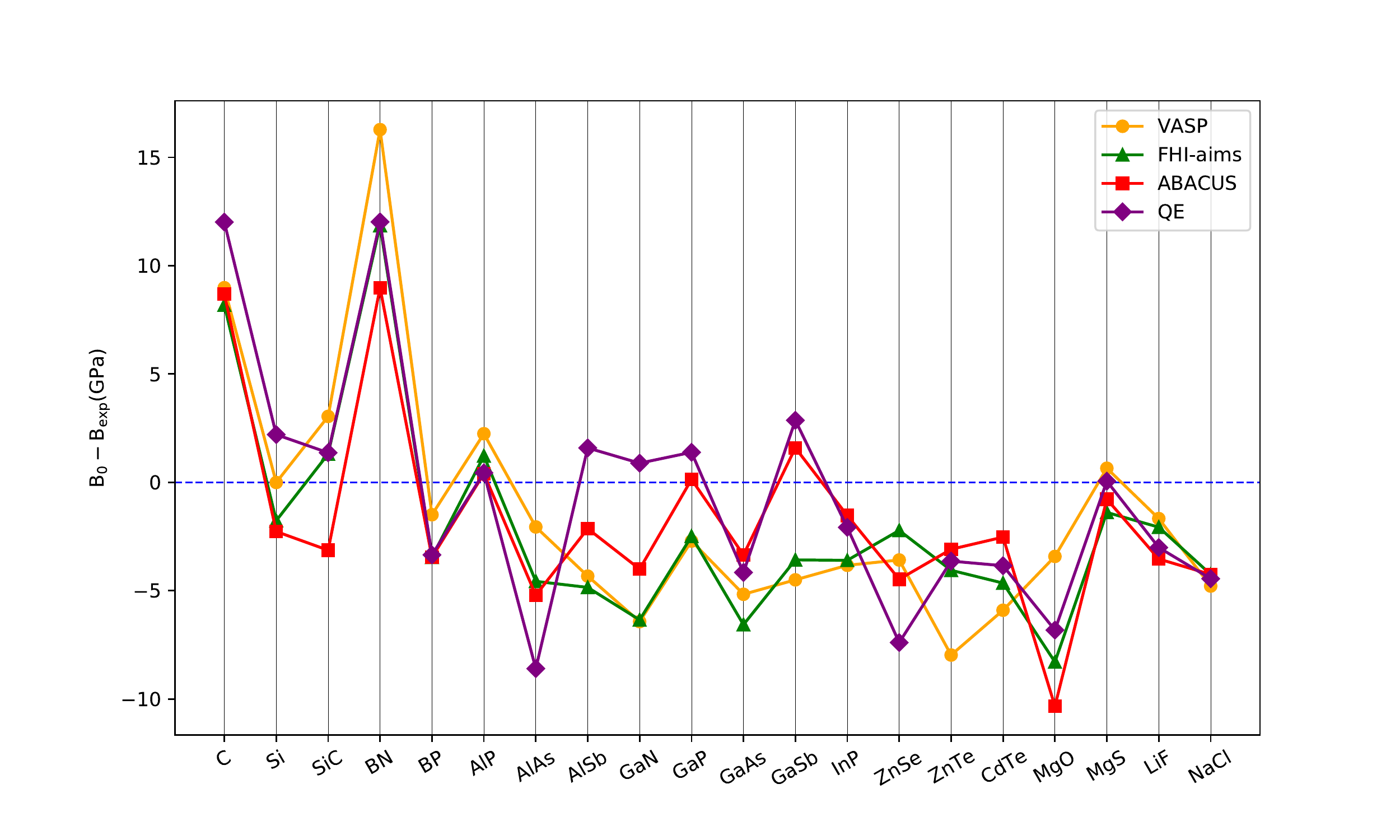}}
    \caption{(a) The pairwise MADs between the computed HSE bulk moduli (in GPa) by the four codes, and between
    the computational and experimental values. 
    The colormap is such that the darker the color, the greater the MADs.
    (b) Deviations between the four sets of HSE bulk moduli and the experimental value for individual
    compounds. The computational bulk moduli are corrected for the ZPM effects in both (a) and (b). }
\end{figure}

We then perform a similar analysis of the bulk moduli for the same set of compounds, based on the EOS curves generated by the four
codes. In Figure~\ref{fig:bulkmoduli_a}, we present
the pair-wise MAD matrix for the experimental and four sets of computed HSE bulk moduli. Again the ZPM corrections to the computed bulk moduli
are included, using the values reported in Ref.~\citenum{Zhang2018b}. From Figure~\ref{fig:bulkmoduli_a}, we see that the computed results also fall into
two groups, with the all-electron FHI-aims and PAW-based VASP results coming closer, and so do the NCPP-based ABACUS and QE results.
Between the two groups, we see that the MAD between the two sets of NAO-based results is smaller than that between the PW-based results. 
Moreover, Figure~\ref{fig:bulkmoduli_a} also reveals that the all-electron/PAW based HSE results show larger deviation from experimental results
than the NCPP-based results, similar to the lattice constant case. In Figure~\ref{fig:bulkmoduli_b}, we further present the deviations of the four sets of computed HSE bulk moduli from the
experimental values for individual materials. The actual values of the HSE bulk moduli as calculated by the four codes, as well as the 
experimental ones, are presented in Table~S3 of SI, for completeness.
Here, although the general better agreements between VASP and FHI-aims results, and between
the ABACUS and QE results are observable, exceptions do exist. For example, for ZnTe and MgO, the VASP results deviate noticeably from the other three
results, whereas the same occurs for the ABACUS results in case of SiC. Despite these discrepancy, some general features can be drawn from this
benchmark study. For example, all four codes predict that the HSE bulk moduli of C and BN are appreciably larger than the experimental value, pointing to
an intrinsic accuracy limitation of HSE for these hard materials.


\subsection{Band gaps}
\begin{figure}[htbp]
    \subfloat[\label{Fig:bandgap_a}]{\includegraphics[scale=0.45]{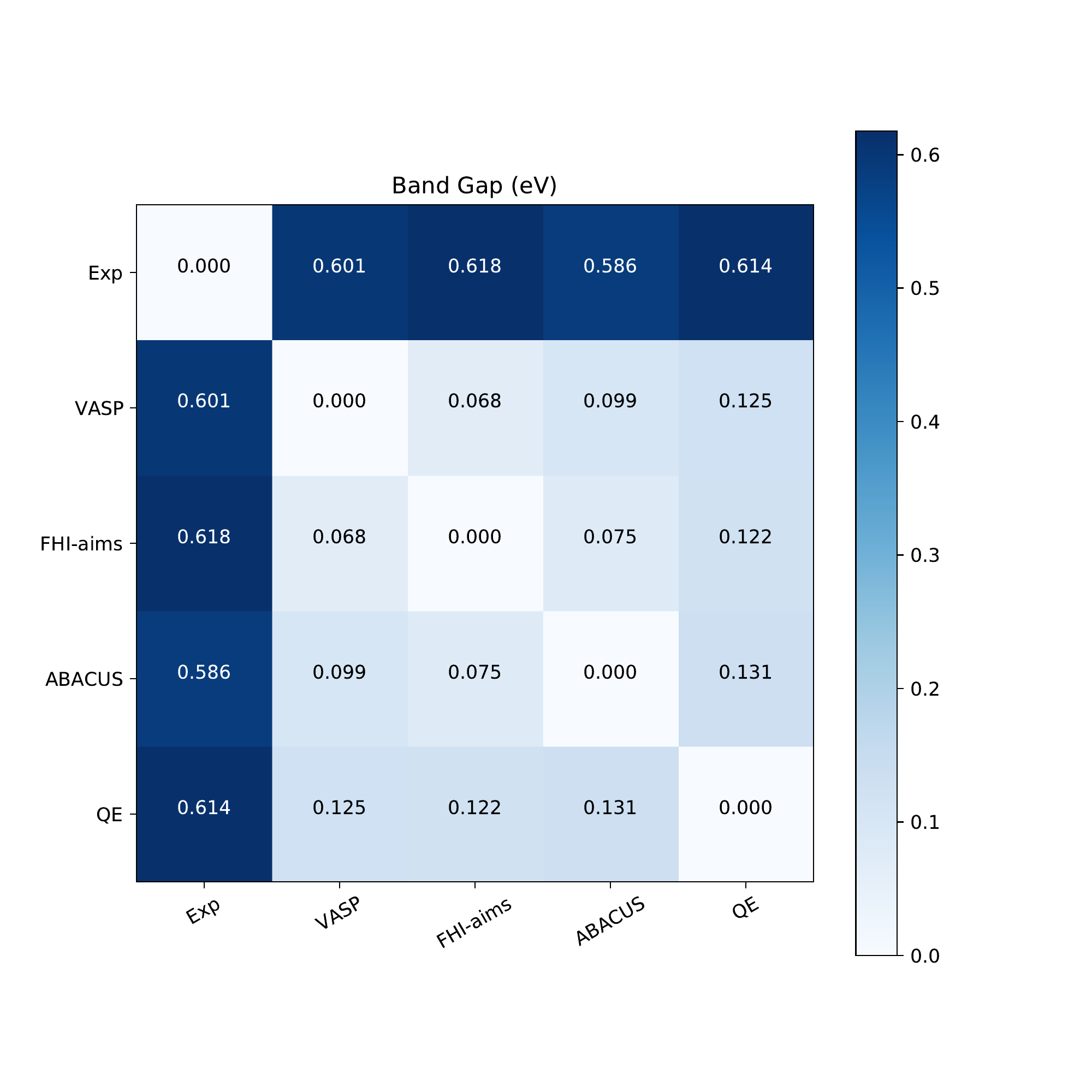}}
    \\
    \subfloat[\label{Fig:bandgap_b}]{\includegraphics[scale=0.45]{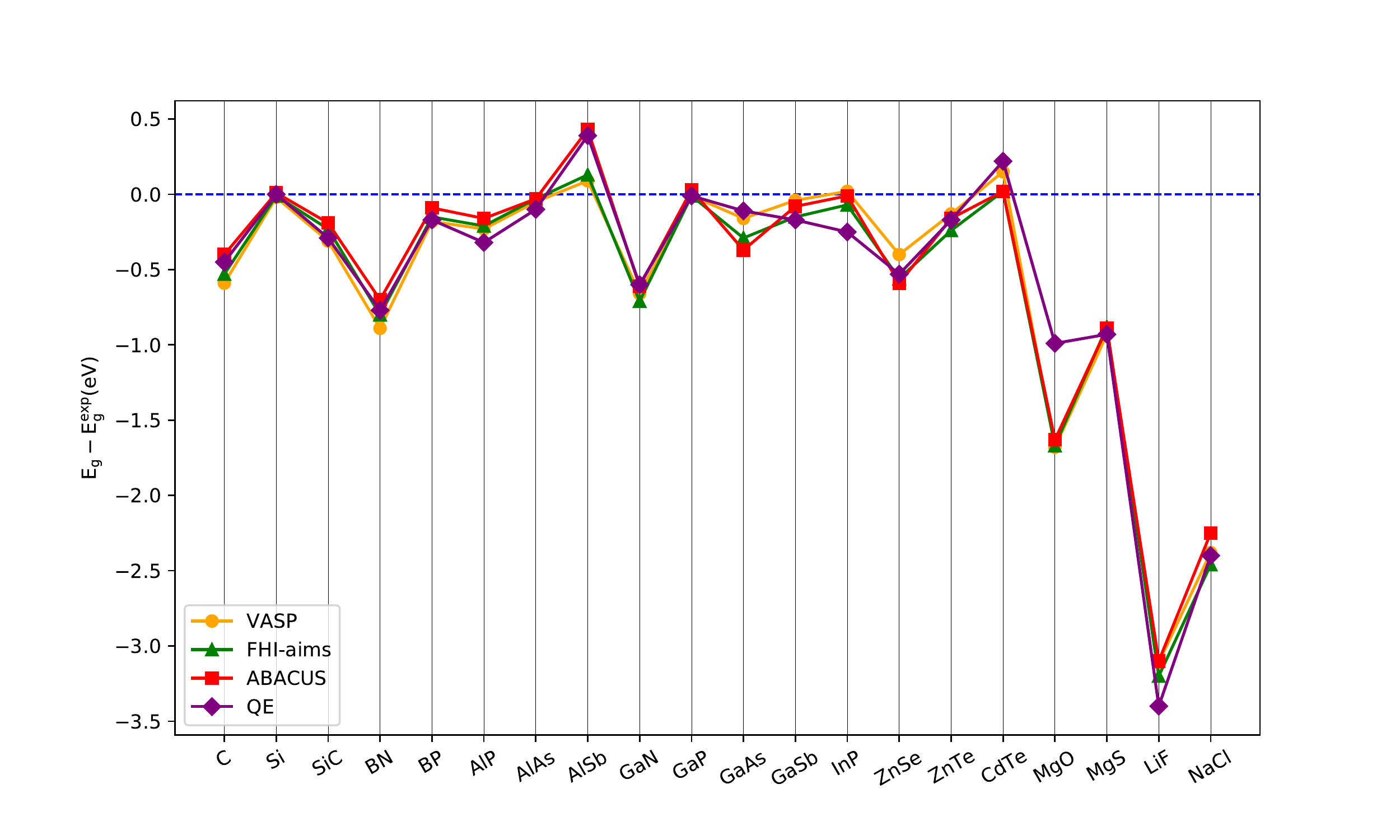}}
    \caption{
    (a) The pairwise MADs between the calculated HSE band gaps (in eV) by the four codes, and between
    the computational and experimental gap results. 
    (b) Deviations between the four sets of HSE band gaps and the experimental value for individual compounds.
    The ZPM contributions to the experimental band gaps are removed here. 
    }
\end{figure}

One important success of the HSE functional is that it significantly improves over local and
semi-local functionals for describing the band gaps of semiconductors and insulators, 
thanks to the HFX component and the generalized KS framework. Thus,  
the reproducibility of the calculated HSE band gaps is also an important aspect to check, in addition to the EOS-based properties. 
In Figure~\ref{Fig:bandgap_a}, we present the MAD matrix for the band gaps, including the experimental and four sets
of HSE results. The actual values of the HSE band gaps, as given by the four codes, as well as the experimental results are presented in 
Table~S4 of the SI. Here, the ZPM contributions to the experimental band gap values are discounted, to facilitate their comparison to 
computational results (see SI for further details). 
Here, we clearly see that the MADs between the experimental and computational results are one order of magnitude larger than
those between the computational results. 
This means that, as far as the band gap concerned, the numerical precision of the major HSE implementation is sufficiently 
high for unambiguously assessing the intrinsic accuracy of the HSE functional. Among the four sets of computational results, the MAD between VASP and FHI-aims results is the smallest, followed by that between FHI-aims and ABACUS. Comparatively, the QE results show larger deviations from those of
the other three codes. The closer agreement between the NCPP-based ABACUS and QE results for EOS-derived quantities does not hold anymore for band
gaps. Figure~\ref{Fig:bandgap_b} further shows the deviations of the four sets of calculated HSE band gaps from the experimental values for 
individual compounds. Here, one can see that the computational results agree with one another well in general, making them overall clearly separated from the
experimental ones. However, there are a few compounds where larger discrepancies are noticeable. These are MgO where the HSE gap by QE shows a deviation
as large as 0.5 eV from the results by other codes, followed by InP where a 0.2 eV discrepancy is noticeable. For AlSb, the VASP and FHI-aims
gap values show good agreement with experiment, whereas ABACUS and QE results overestimate by the band gap by about 0.4 eV.


\section{Conclusions}

In this work, we examined the reproducibility of HSE calculations based on four different implementations in the VASP, FHI-aims, QE, and ABACUS codes, 
in terms of a test set comprising 20 semiconductors and insulators. These implementations are representative ones as different procedures for treating
the core-valence interactions (all-electron, PAW, and NCPP schemes) and different basis sets (PWs and atomic orbitals), with their different
combinations, are used. 
The quantities that we have checked include the $\Delta$ metric as introduced
in Ref.~\citenum{Lejaeghere.2014}, as well as the lattice constants, bulk moduli, and
the band gaps. We found that, for band gaps, the deviations between the four different sets of computational results
are nearly one order of magnitude smaller than the difference between the computational and experimental results. For the EOS-derived properties, e.g., the lattice constants and bulk moduli, the differences between computational and experimental results and those between the
computational results themselves, are at the same order of magnitude. On the one hand, this shows the rather good performance of the HSE functional for
describing the EOS-derived properties; on the other hand, this also means that the numerical precision of HSE implementation still needs to be improved to
unambiguously assess its intrinsic accuracy with respect to experimental reference data. Among the computational results, it seems that the prescription for
describing the ion-cores plays a significant role in determining the numerical precision of the HSE implementations. Specifically, the all-electron (FHI-aims) 
and PAW-based (VASP) implementations show better agreement with each other, and so does the NCPP-based implementations (QE and ABACUS), whereas the 
deviations between these two sets of results are noticeably larger. Nowadays, most NCPP-based HSE calculations are performed using the PBE pseudopotentials. Our work indicates that it is not ideal if high-precision results are aimed for. In principle, one can reduce the core radius and increase the energy cutoffs for NCPP-based calculations to approach the precision of all-electron calculations. However, the computational cost will get exceedingly high to do so. Designing improved NCPPs for HSE and other types of HDF calculations would be a better strategy for improving the quality of NCPP-based HDF calculations. Yang et al. have shown that using the PBE0 pseudopotentials in PBE0 calculations improves the agreement with all-electron calculations for structural parameters and band gaps of some small molecules and simple solids \cite{Yang.2018}. We expect that this will 
also holds for the HSE functional and the benchmark set presented in this work.

As HDF calculations get increasingly popular in computational materials sciences, the reproducibility of this type of calculations becomes a prominent 
issue. Our benchmark calculations in this work only cover a limited set of codes and materials, yet interesting features and trends have been revealed.
We hope that this work will stimulate more comprehensive investigations along this line.

\begin{suppinfo}

The Supporting Information is available free of charge .
\begin{itemize}
  \item Description of the computational parameters used in ABACUS calculations, and tables containing the computational and experimental values of lattice constants, bulk moduli, and band gaps for 20 solids.
\end{itemize}

\end{suppinfo}

\section{Data availability}
All output files of the calculations using VASP, FHI-aims, and QE reported in this work have been uploaded to \href{https://nomad-lab.eu/}{the NOMAD repository}, and can be found, respectively, under the
links \\
\href{https://nomad-lab.eu/prod/v1/gui/user/uploads/upload/id/3WA4ETfQSEecuOOPo6kMSQ}{https://nomad-lab.eu/prod/v1/gui/user/uploads/upload/id/3WA4ETfQSEecuOOPo6kMSQ}, \\
\href{https://nomad-lab.eu/prod/v1/gui/user/uploads/upload/id/CxYC8UVYR7Wv7EP1FXhgKw}{https://nomad-lab.eu/prod/v1/gui/user/uploads/upload/id/CxYC8UVYR7Wv7EP1FXhgKw}, \\ \href{https://nomad-lab.eu/prod/v1/gui/user/uploads/upload/id/1I\_V\_-9lR_WKrXiTfsF0Fw}{https://nomad-lab.eu/prod/v1/gui/user/uploads/upload/id/1I\_V\_-9lR\_WKrXiTfsF0Fw}. \\
The input and output files of ABACUS calculations are available from the authors upon reasonable request.

\begin{acknowledgement}
This work is supported by National Natural Science Foundation of China (Grant Nos. 12134012, 11774327, 11874335, 12188101).
The author thanks Dr. Wenshuai Zhang for his help in $\Delta$-value calculations. The numerical calculations have been done on the USTC HPC facilities.

\end{acknowledgement}

\bibliography{hse-test}

%

\end{document}


\renewcommand{\thepage}{S\arabic{page}}
\renewcommand{\thefigure}{S\arabic{figure}}
\renewcommand{\thetable}{S\arabic{table}}
\renewcommand{\theequation}{S\arabic{equation}}

Here,  we present more details about the the computational parameters used in ABACUS calculations, so that the results obtained
by ABACUS can be reproduced by interested readers using these settings. To better explain the meanings of the computational parameters involved,
we briefly introduce the localized RI (LRI) scheme\cite{ihrig2015accurate} on which the hybrid density functional (HDF) implementation of ABACUS \cite{lin2020accuracy,Lin2021} is based.

Within the linear combination of atomic orbitals (LCAO) framework, the elements of the Hartree-Fock exchange (HFX) matrix can be formally calculated as \cite{Lin2021},
\begin{equation}
    \Sigma_{Ii,Jj}^\text{HFX}=\sum_{K,L}\sum_{k\in K,l\in L}(\bmphi_{Ii}\bmphi_{Kk}|\bmphi_{Jj}\bmphi_{Ll})D_{Kk,Ll}\label{eq:hfx}
\end{equation}
where the abbreviation
\begin{equation}
    (f|g)\overset{\text{def}}{=}\int\int{f(\bfr)v(\bfr-\bfrp)g(\bfr)d\bfr d\bfrp}\label{eq:sc}
\end{equation}
 denotes the interaction integral between two functions $f(\bfr)$ and $g(\bfr)$ via the bare or screened Coulomb potential
 $v(\bfr-\bfrp)$. The capital letters ($I$, $J$, $K$, $L$) indicate atoms, while the lowercase letters ($i$, $j$, $k$, $l$) indicate atomic orbitals (AOs) centering on these atoms. $D_{Kk,Ll}$ is the density matrix in real space. Within the LRI scheme, the product of two numerical atomic orbitals (NAOs) centering on two atoms is approximately expanded by a set of ABFs:
\begin{equation}
    \bmphi_{Ii}(\bfr)\bmphi_{Kk}(\bfr)\approx\sum_{A=\{I,K\}}\sum_{\alpha\in A}C^{A\alpha}_{Ii,Kk}P_{A\alpha}(\bfr) \label{eq:abfs}
\end{equation}
where $C^{A\alpha}_{Ii,Kk}$ are the two-center, three-index expansion coefficients. The Greek letters ($\alpha$, $\beta$) are indices of ABFs. In ABACUS, there are two types of ABFs called ``on-site'' and ``off-site'' basis (also called \textit{optimized} ABFs in Ref.~\citenum{lin2020accuracy}). The former ones are generated from the pair products of AOs on the same atom and a principal component analysis (PCA) is used to remove the linear dependence, while the latter ones, as a complement to
the the ``on-site'' ones \cite{lin2020accuracy}, are generated to best represent the products of AOs from two different atoms. The two parameters \texttt{exx\_pca\_threshold} and \texttt{off-site basis} in Table.~\ref{table:hfx-settings} denote PCA threshold and the ``off-site'' ABFs, respectively.

Then the HFX in eq \refeq{eq:hfx} is naturally split into four pieces, using Eq.~\ref{eq:abfs},
\begin{equation}
\begin{split}
    \Sigma_{Ii,Jj}^{\rm HFX} &= \sum_{K,L}\sum_{k\in K,l\in L}(\sum_{\alpha\in I,\beta\in J}{C^{I\alpha}_{Ii,Kk}V_{I\alpha,J\beta}C^{J\beta}_{Jj,Ll}} \\
            &+ \sum_{\alpha\in I,\beta\in L}{C^{I\alpha}_{Ii,Kk}V_{I\alpha,L\beta}C^{L\beta}_{Jj,Ll}} \\
            &+ \sum_{\alpha\in K,\beta\in J}{C^{K\alpha}_{Ii,Kk}V_{K\alpha,J\beta}C^{J\beta}_{Jj,Ll}} \\
            &+ \sum_{\alpha\in K,\beta\in L}{C^{K\alpha}_{Ii,Kk}V_{K\alpha,L\beta}C^{L\beta}_{Jj,Ll}}
    )D_{Kk,Ll}
\end{split}
\label{eq:hfx-abfs}
\end{equation}
where
\begin{equation}
\begin{split}
    V_{A\alpha, B\beta}&=(P_{A\alpha}|P_{B\beta}) \\
                    &= \bracketw{P_{A\alpha}}{Q_{B\beta}}
\end{split}\label{eq:v}
\end{equation}
with $A\in\{I,K\}$ and $B\in\{J,L\}$ are (screened) Coulomb integrals between two ABFs and
$Q_{B\beta}(\bfr)=\int{v(\bfr-\bfrp)P_{B\beta}(\bfrp)d\bfrp}$. For screened HDFs such as HSE, a short-ranged (screened) Coulomb interaction $v(\bfr-\bfrp)=\text{erfc}(\omega|\bfr-\bfrp|)/|\bfr-\bfrp|$ is employed, so $Q_{\beta}(\bfr)$ is short-ranged. Hence, ABACUS introduce a cut-off radius for $Q_{\beta}(\bfr)$, and \texttt{exx\_ccp\_rmesh\_times} is used to control the size of this radius.

Because of the locality of the NAOs and, in the case of insulators, the rapidly decaying behavior of the density matrix in real space, most of elements of matrices in \refeq{eq:hfx-abfs} (i.e. expansion coefficients matrix $C$, Coulomb matrix $V$ and density matrix $D$) are extremely small or even strictly zero. Hence, a pre-screening procedure is invoked to filter out these negligibly small contributions, and \texttt{exx\_c\_threshold}, \texttt{exx\_v\_threshold} and \texttt{exx\_dm\_threshold} in Table.~\ref{table:hfx-settings} are pre-screening thresholds of C, V and D, respectively. 
In addition, ABACUS employs an screening algorithm based on the Cauchy-Schwarz inequality to further neglect small contributions from the whole blocks (CVCD) in eq \refeq{eq:hfx-abfs}, and \texttt{exx\_schwarz\_threshold} and \texttt{exx\_cauchy\_threshold} denote two different upper bounds of the elements of blocks, respectively. 

\begin{table}[H]
    \centering
    \caption{Settings for the numerical and screening parameters used in HSE calculations of ABACUS done performed in this work.} \label{table:hfx-settings}
    \begin{tabular}{ c|c }
         \hline
            Key & Value  \\
        \hline
        \hline
            \texttt{off-site basis} & [4$s$ 3$p$ 2$d$ 1$f$] \\
            \texttt{exx\_pca\_threshold} & $10^{-4}$  \\
            \texttt{exx\_c\_threshold} & $5\times 10^{-5}$  \\
            \texttt{exx\_v\_threshold} & 0 \\
            \texttt{exx\_dm\_threshold} & $5\times 10^{-4}$  \\
            \texttt{exx\_schwarz\_threshold} & $5\times 10^{-5}$  \\
            \texttt{exx\_cauchy\_threshold} & $10^{-8}$  \\
            \texttt{exx\_ccp\_rmesh\_times} & 1.5  \\
         \hline
         \hline
    \end{tabular}
\end{table}

Table~\ref{table:lat} and \ref{table:B} present the actual values of the lattice constants and bulk moduli of the 20 solids as computed by the four codes, 
as well as those from experiment. 
The ``Ref'' column contains independent results taken from Ref.~\citenum{Zhang2018b}, which are also obtained using FHI-aims. In particular,
the zero-point motion (ZPM) corrections to the lattice constants and bulk moduli are estimated at the HSE level in Ref.~\citenum{Zhang2018b}, 
and presented in the ``ZPM'' columns of Table~\ref{table:lat} and \ref{table:B}, respectively. These ``ZPM'' results are used to estimate the
MADs between the computational and experimental results in the main text. In their calculations, Zhang \textit{et al.} also used ``\emph{tight}'' setting in FHI-aims, but with different convergence criteria, i.e., $10^{-5}$ electrons for the electron density, $10^{-3}$ eV for the eigenvalues, and $10^{-6}$ eV for the total energy of the system\cite{Zhang2018b}, whereas only the electron density criterion with a threshold of   $10^{-6}$ is used in our work. 
This leads to small but negligible differences. Indeed, the mean absolute deviations (MADs) between results in Ref.~\citenum{Zhang2018b} and 
ours are only 0.001 \AA~for lattice constants, and 0.8 GPa for bulk moduli. Such differences have negligible influence on
the analysis presented in Figure~2 and 3 in the main text. 

\begin{table}[H]
    \centering
    \caption{Computational and experimental lattice constants (in \AA) of 20 fcc solids. The computational results are original and not yet corrected for ZPM effects. Independent computational results using FHI-aims (column ``Ref'') are also listed for comparison.
    (*: the sources of experimental values are indicated explicitly
    below the table.)} \label{table:lat}
    \begin{tabular}{c c c c c c c c}
        \hline
            & VASP  & QE & ABACUS & FHI-aims & Ref.\cite{Zhang2018b} & ZPM\cite{Zhang2018b}  & $\rm Exp^*$ \\
        \hline
        \hline
C& 3.549  & 3.541  & 3.546  &3.548 &3.549 &0.012 & 3.567  \\
Si& 5.436  & 5.444  & 5.456  &5.445 &5.444 &0.008 & 5.430 \\
SiC& 4.348  & 4.351  & 4.368  &4.353 &4.353 &0.010 & 4.358\\
BN& 3.599  & 3.594  & 3.602  &3.600 &3.601 &0.013 & 3.607 \\
BP& 4.521  & 4.517  & 4.529  &4.529 &4.530 &0.012 & 4.538 \\
AlP& 5.472  & 5.483  & 5.492  &5.480 &5.482 &0.009 & 5.463\\
AlAs& 5.686  & 5.709  & 5.715  &5.695 &5.694 &0.004 & 5.652  \\
AlSb& 6.191  & 6.116  & 6.136  &6.185 &6.184 &0.007 & 6.128  \\
GaN& 4.494  & 4.481  & 4.503  &4.503 &4.504 &0.009 & 4.520  \\
GaP& 5.461  & 5.420  & 5.434  &5.469 &5.467 &0.008 & 5.442  \\
GaAs& 5.686  & 5.653  & 5.670  &5.695 &5.692 &0.005 & 5.641  \\
GaSb& 6.156  & 6.038  & 6.069  &6.155 &6.152 &0.006 & 6.082  \\
InP& 5.903  & 5.852  & 5.878  &5.909 &5.907 &0.007 & 5.861  \\
ZnSe& 5.709  & 5.712  & 5.741  &5.699 &5.700 &0.006 & 5.668  \\
ZnTe& 6.161  & 6.110  & 6.141  &6.149 &6.152 &0.007 & 6.104  \\
CdTe& 6.583  & 6.521  & 6.539  &6.571 &6.574 &0.005 & 6.480  \\
MgO& 4.195  & 4.211  & 4.232  &4.208 &4.208 &0.016 & 4.207  \\
MgS& 5.184  & 5.200  & 5.218  &5.207 &5.207 &0.014 & 5.202  \\
LiF& 4.040  & 4.019  & 4.020  &4.002 &4.003 &0.037 & 4.010  \\
NaCl& 5.660  & 5.648  & 5.662  &5.643 &5.645 &0.022 & 5.595  \\
        \hline
        \hline
    \end{tabular}
\end{table}

* C\cite{stoupin2010thermal}, Si\cite{reeber1996thermal}, SiC\cite{staroverov2004tests,staroverov2008erratum}, BN\cite{madelung2004semiconductors}, BP\cite{madelung2004semiconductors}, AlP\cite{vurgaftman2001band}, AlAs\cite{vurgaftman2001band}, AlSb\cite{vurgaftman2001band}, GaN\cite{haas2009calculation}, GaP\cite{vurgaftman2001band}, GaAs\cite{vurgaftman2001band}, GaSb\cite{vurgaftman2001band}, 
InP\cite{vurgaftman2001band}, ZnSe\cite{madelung2004semiconductors,ley1974total}, ZnTe\cite{ley1974total},
CdTe\cite{ley1974total}, MgO\cite{staroverov2004tests,staroverov2008erratum}, MgS\cite{peiris1994compression}, 
LiF\cite{staroverov2004tests,staroverov2008erratum}, NaCl\cite{staroverov2004tests,staroverov2008erratum}

\begin{table}[H]
    \centering
    \caption{Computational and experimental bulk moduli (in GPa) of 20 fcc solids. The computational results are original and not yet corrected for ZPM effects. Independent computational results obtained using FHI-aims (column ``Ref'') are also listed for comparison.  (*: the references where
    the experimental values are taken are given
    below the table.)} \label{table:B}
    \begin{tabular}{ c c c c c c c c}
        \hline
            & VASP  & QE & ABACUS & FHI-aims & Ref\cite{Zhang2018b}  & ZPM\cite{Zhang2018b} & $\rm Exp^*$ \\
        \hline
        \hline
C& 467.9  & 470.9  & 467.6  & 467.1  &468.8 &-15.90 & 443.0  \\
Si& 99.4  & 101.6  & 97.1  & 97.6  &97.6 &-0.20 & 99.2  \\
SiC& 231.5  & 229.9  & 225.4  & 229.8  &228.9 &-3.50 & 225.0 \\
BN& 404.1  & 399.8  & 396.8  & 399.7  &399.6 &-9.80 & 378.0  \\
BP& 174.8  & 173.0  & 172.8  & 172.8  &173.0 &-3.30 & 173.0  \\
AlP& 91.1  & 89.3  & 89.3  & 90.1  &94.5 &-2.90 & 86.0  \\
AlAs& 76.4  & 69.9  & 73.3  & 73.9  &74.5 &-0.40 & 78.1 \\
AlSb& 55.7  & 61.6  & 57.9  & 55.2  &55.2 &-0.70 & 59.3 \\
GaN& 197.7  & 205.0  & 200.1  & 197.7  &197.4 &-4.10 & 200.0 \\
GaP& 87.5  & 91.6  & 90.3  & 87.7  &87.9 &-1.50 & 88.7  \\
GaAs& 72.6  & 73.6  & 74.4  & 71.2  &70.9 &-0.90 & 76.9 \\
GaSb& 52.1  & 59.5  & 58.2  & 53.0  &53.0 &-0.30 & 56.3 \\
InP& 69.7  & 71.4  & 72.0  & 69.9  &69.4 &-1.00 & 72.5  \\
ZnSe& 62.5  & 58.7  & 61.6  & 63.9  &63.3 &-1.40 & 64.7 \\
ZnTe& 45.1  & 49.5  & 50.0  & 49.0  &48.9 &-0.30 & 52.8 \\
CdTe& 39.3  & 41.4  & 42.7  & 40.5  &40.1 &-0.70 & 44.5 \\
MgO& 169.9  & 166.5  & 163.0  & 165.0  &165.0 &-4.50 & 168.8 \\
MgS& 80.5  & 80.0  & 79.1  & 78.5  &78.8 &-1.00 & 78.9  \\
LiF& 73.7  & 72.4  & 71.9  & 73.3  &77.9 &-5.60 & 69.8  \\
NaCl& 24.9  & 25.3  & 25.4  & 25.5  &25.3 &-1.20 & 28.5 \\
        \hline
        \hline
    \end{tabular}
\end{table}

* C\cite{staroverov2004tests,staroverov2008erratum}, Si(77 K)\cite{hall1967electronic}, SiC\cite{carnahan1968elastic}, BN\cite{aguado2006prediction}, BP\cite{wettling1984elastic}, AlP\cite{tan2012elastic},
        AlAs\cite{tan2012elastic,harrison2012electronic}, AlSb\cite{tan2012elastic,harrison2012electronic}, GaN\cite{kisielowski1996strain}, GaP\cite{tan2012elastic,harrison2012electronic}, 
        GaAs\cite{cottam1973elastic}, GaSb\cite{tan2012elastic,harrison2012electronic}, InP\cite{tan2012elastic,harrison2012electronic}, ZnSe\cite{collins1980thermal}, ZnTe\cite{collins1980thermal},
        CdTe\cite{collins1980thermal}, MgO\cite{cohen1976modified}, MgS\cite{peiris1994compression}, LiF\cite{briscoe1957elastic}, NaCl\cite{spetzler1972equation}

Table.~\ref{table:gap} present the actual computed band gaps of the 20 solids as given by the four codes. Original experimental values ($E_g(T=0)$) are summarized in the last columns. The ZPM contributions estimated from extrapolation of temperature dependence of the gap, or from isotopic mass derivatives of the gap or from theoretical calculations are listed in the column ``ZPM" for each material. Theoretical estimated values for ZPM contributions
are adopted for those materials
where experimental results are not available in the literature. Corrected experimental band gaps  $E_g^{\text{corr}}=E_g(T=0)-\Delta E_g^{\text{ZPM}}$ are used for the MAD analysis presented in Figure~4 of the main text.

\begin{table}[H]
    \centering
    \caption{Computational and experimental band gaps (in eV) of 20 fcc solids.  (*: the references where the experimental values and theoretical estimates of ZPM  are taken are given below the table. Theoretical estimates are noted by ``$t$'' in parentheses.)} \label{table:gap}
    \begin{tabular}{ c c c c c c c }
        \hline
            & VASP & QE & ABACUS & FHI-aims & ZPM$^*$ & Exp\cite{Martienssen1996Landolt,madelung2004semiconductors} \\
        \hline
        \hline
C&5.26 &5.40 &5.45 &5.32 & -0.370 &5.48 \\
Si&1.21 &1.24 &1.24 &1.22 & -0.064 &1.17 \\
SiC&2.26 &2.27 &2.37 &2.34 & -0.145 &2.42 \\
BN&5.78 &5.90 &5.96 &5.86 & -0.262 &6.40 \\
BP&2.03 &2.03 &2.11 &2.06 & -0.106 &2.10 \\
AlP&2.30 &2.21 &2.38 &2.32 & -0.023 &2.51 \\
AlAs&2.09 &2.04 &2.11 &2.11 & -0.039 &2.10 \\
AlSb&1.73 &2.03 &2.07 &1.77 & -0.039 &1.60 \\
GaN&2.82 &2.87 &2.87 &2.76 & -0.173 & 3.30 \\
GaP&2.33 &2.33 &2.37 &2.34 & -0.085 & 2.26 \\
GaAs&1.32 &1.37 &1.11 &1.19 & -0.060 &1.42 \\
GaSb&0.81 &0.68 &0.77 &0.70 & -0.040 & 0.81 \\
InP&1.41 &1.14 &1.38 &1.32 & -0.054 & 1.34 \\
ZnSe&2.36 &2.23 &2.17 &2.19 & -0.055 &2.70 \\
ZnTe&2.17 &2.13 &2.14 &2.06 & -0.040 &2.26 \\
CdTe&1.59 &1.67 &1.47 &1.47 & -0.016 &1.43 \\
MgO&6.30 &6.99 &6.36 &6.32 & -0.154 &7.83 \\
MgS&3.75 &3.75 &3.79 &3.79 & -0.179 & 4.50 \\
LiF&11.37 &11.09 &11.38 &11.29 & -0.281 & 14.20 \\
NaCl&6.41 &6.39 &6.54 &6.33 &-0.290& 8.50 \\
        \hline
        \hline
    \end{tabular}
\end{table}
* C\cite{Cardona.2005, Ren.2021}, Si\cite{Cardona.2005, Ren.2021}, SiC\cite{Cardona.2005, Ren.2021}, BN($t$)\cite{Antonius.2015}, BP($t$)\cite{Shang.2021}, AlP\cite{Cardona.2005, Ren.2021},
        AlAs\cite{Cardona.2005, Ren.2021}, AlSb\cite{Cardona.2005, Ren.2021}, GaN\cite{Cardona.2005, Ren.2021}, GaP\cite{Cardona.2005, Ren.2021}, 
        GaAs\cite{Roland.2002, Miglio.2020}, GaSb\cite{Roland.2002}, InP\cite{Roland.2002}, ZnSe\cite{Roland.2002, Miglio.2020}, ZnTe\cite{Roland.2002, Miglio.2020},
        CdTe\cite{Roland.2002, Miglio.2020}, MgO($t$)\cite{Antonius.2015}, MgS($t$)\cite{Shang.2021}, LiF($t$)\cite{Antonius.2015}, NaCl($t$)\cite{Shang.2021}

\bibliography{hse-test}

%% file: newcommands.tex
\newcommand{\bfk}{ {\bf k}} 

